\documentclass[%
twocolumn,
 amsmath,amssymb,
 aps,authoryear
]{aastex631}
\usepackage{graphicx}
\usepackage{ulem}
\usepackage{epstopdf}
\usepackage{amsmath}
\usepackage{amssymb}
\usepackage{mathtools}
\usepackage{mathrsfs}
\usepackage{comment}
\usepackage{cancel}
\usepackage{bm}
\usepackage{color}
\usepackage{textcomp}
\usepackage{gensymb}
\usepackage{hyperref}
\newcommand{\dash}{\text{-}}

\newcommand{\kmps}{\text{km s}^{-1}}          

\newcommand{\curl}{\nabla\times}

\newcommand{\ddt}[1]{\frac{\partial #1}{\partial t}}

\renewcommand{\S}{\hbox{$\cal S$}}

\newcommand{\M}{\hbox{$\cal M$}}
\newcommand{\E}{\hbox{$\mathcal{E}$}}
\newcommand{\Rrp}{\mathcal{R}_{r\phi}}

\begin{document}

\title{The effect of turbulence on the angular momentum of the solar wind}

\author[0000-0002-7174-6948]{Rohit Chhiber}
\affiliation{Bartol Research Institute and Department of Physics and Astronomy, University of Delaware, Newark, DE 19716, USA}
\affiliation{Heliophysics Science Division, NASA Goddard Space Flight Center, Greenbelt, MD 20771, USA}
\email{rohit.chhiber@nasa.gov}

\author[0000-0002-0209-152X]{Arcadi V. Usmanov}
\affiliation{Bartol Research Institute and Department of Physics and Astronomy, University of Delaware, Newark, DE 19716, USA}
\affiliation{Heliophysics Science Division, NASA Goddard Space Flight Center, Greenbelt, MD 20771, USA}

\author[0000-0001-7224-6024]{William H. Matthaeus}
\affiliation{Bartol Research Institute and Department of Physics and Astronomy, University of Delaware, Newark, DE 19716, USA}

\author[0000-0003-4168-590X]{Francesco Pecora}
\affiliation{Bartol Research Institute and Department of Physics and Astronomy, University of Delaware, Newark, DE 19716, USA}

\begin{abstract}
The transfer of a star's angular momentum to its atmosphere is a topic of considerable and wide-ranging interest in astrophysics. This letter considers the effect of kinetic and magnetic turbulence on the solar wind's angular momentum. The effects are quantified in a theoretical framework that employs Reynolds-averaged mean field magnetohydrodynamics, allowing for fluctuations of arbitrary amplitude. The model is restricted to the solar equatorial (\(r\dash\phi\)) plane with axial symmetry, which 
permits the effect of turbulence to be expressed in analytical form as a modification to the classic \cite{weber1967ApJ148} theory, dependent on the \(r,\phi\) shear component of the Reynolds stress tensor. A solar wind simulation with turbulence transport modeling and Parker Solar Probe observations at the Alfv\'en surface are employed to
quantify 
this turbulent modification to the solar wind's angular momentum, which is 
found to be 
\(\sim3\% ~\dash~10\%\) and tends to be negative. Implications for solar and stellar rotational evolution are discussed. 
\end{abstract}


\section{Introduction}

Angular momentum is a fundamental property
of stars, planetary systems, galaxies,
and other astrophysical objects, 
representing a significant factor in establishing structure and controlling evolution. 
In our own cosmic environment, only a small fraction of the total angular momentum resides in the Sun \citep[e.g.,][]{Kuiper1951PNAS},
but its budget and loss rate are key parameters
in controlling evolution of the Sun, its 
dynamo activity, and how it relates to and influences the heliosphere. 
In particular, a quantity of central interest 
is the rate of loss of solar angular momentum
associated mainly with its outward transport by the solar wind. 
In this Letter we quantify the 
modifications of the classical treatment of 
solar angular momentum loss due to  turbulence in the solar wind,
employing analytical treatments, numerical simulation, and spacecraft observations.

\section{Background}\label{sec:backg}

The early work by \cite{weber1967ApJ148} (henceforth WD67) used the ideal magnetohydrodynamic (MHD) equations to describe a solar wind restricted to the equatorial plane under the assumption of axial symmetry, and derived the following expression for the constant angular momentum per unit mass, in a steady state:
\begin{equation}
    \mathscr{L} = r{u}_\phi - \frac{r{B}_r {B}_\phi}{4\pi\rho {u}_r}, \label{eq:WD_L}
\end{equation}
with \(d\mathscr{L}/dr = 0\). Here \(r\) is heliocentric distance, \(u_\phi\) and \(u_r\) are azimuthal and radial components, respectively, of the solar wind velocity in a spherical coordinate system. \(B_r\) and \(B_\phi\) are the corresponding components of the magnetic field, and \(\rho\) is the mass density of the solar wind. \(\mathscr{L}\) contains the gas contribution in the first term and the contribution of magnetic stresses in the second term of Eq. \eqref{eq:WD_L}.  WD67 also derived the result
\begin{equation}
    \mathscr{L} = r_A^2 \Omega,
    \label{eq:WD_L_rigid}
\end{equation}
where \(r_A\) is the Alfv\'en radius \citep[where the wind speed equals the Alfv\'en speed;][]{Cranmer2023SoPh} and \(\Omega\) is the solar rotation rate. Note that \(r_A^2 \Omega\) resembles the angular momentum per unit mass of matter corotating with the Sun as a solid body with radius \(r_A\). The rate of change of total angular momentum of the Sun (\(\mathscr{J}_\odot\)) was derived by assuming that the calculations made for the equatorial plane [i.e., Eq. \eqref{eq:WD_L_rigid}] apply to the entire angular surface and that the mass flux is spherically symmetric:
\begin{equation}
    \dot{\mathscr{J}}_\odot = \frac{2}{3}\Omega r_A^2 \dot{M}_\odot, \label{eq:WD_Jdot}
\end{equation}
where \(\dot{M}_\odot\) is the solar mass loss rate. WD67 estimated a characteristic time for solar angular momentum depletion of  \(\sim7\times 10^9\) years. The angular momentum lost by the Sun to the solar wind can thus be significant over its nuclear life span (\(\sim 10^{10}\) years).

While later refinements included an accounting of viscosity and pressure anisotropy \citep{Weber1970SoPh,Weber1970JGR,reville2020ApJS}, the effect of fluctuations and turbulence on the solar wind's angular momentum has received relatively limited attention. \cite{schubert1968ApJ} and \cite{Hollweg1973JGR} evaluated the effects of MHD waves in a linearized description, finding that fast mode waves may carry a substantial angular momentum while Alfv\'en waves do not. \cite{usmanov2018} considered the (more general) strong turbulence case, and results from their three-dimensional (3D) numerical model demonstrate that the contribution of turbulence to the solar wind's angular momentum can be significant. In this Letter we follow an approach similar to that of \cite{usmanov2018}, but restrict our analysis to the analytically tractable case of a WD67-type equatorial and axially-symmetric wind. The effects of turbulence can then be expressed as modifications to the WD67 formulae (Sec \ref{sec:theory}). Numerical modeling and in-situ observations by the Parker Solar Probe (PSP) spacecraft at the solar wind's Alfv\'en surface will provide quantitative estimates of this turbulent modification (Sec \ref{sec:results}).

\section{Angular momentum of the solar wind from Reynolds-averaged MHD equations with turbulence}\label{sec:theory}
We carry out a straightforward extension of the WD67 approach while accounting for turbulence using the Reynolds-averaging framework. Our starting point is the set of single-fluid ideal MHD equations for the solar wind \citep[e.g.,][]{lamers1999book,usmanov2000global}:
\begin{equation}
\ddt{\rho} + \nabla\cdot(\rho\tilde{\bm{u}}) = 0, \label{eq:rho}
\end{equation}
\begin{equation}
    \rho\ddt{\tilde{\bm{u}}} + \rho\tilde{\bm{u}}\cdot\nabla\tilde{\bm{u}} = -\nabla P 
       -\frac{GM_\odot}{r^2}\hat{\bm{r}} \\
       +\frac{1}{4\pi} (\curl \tilde{\bm{B}})\times \tilde{\bm{B}}, \label{eq:mom}
\end{equation}
\begin{equation}
\ddt{\tilde{\bm{B}}} = \nabla\times(\tilde{\bm{u}}\times\tilde{\bm{B}}).
  \label{eq:ind}
\end{equation}
The independent variables are heliocentric position vector \(\bm{r}\) and time $t$. Dependent variables are velocity \(\tilde{\bm{u}}\), magnetic field \(\tilde{\bm{B}}\), mass density $\rho$, and pressure \(P\). $G$ and $M_{\sun}$ are the gravitational constant and solar mass, respectively. We consider steady-state conditions, assume axial symmetry about the solar rotation axis (\(\partial/\partial\phi =0\)), and restrict our analysis to the solar equatorial plane, neglecting \(\tilde{u}_\theta\) and \(\tilde{B}_\theta\). Then Eq. \eqref{eq:rho} yields \(r^2\rho \tilde u_r= \text{constant}\), while the solenoidal condition \(\nabla\cdot \tilde{\bm{B}}=0\) becomes \(r^2\tilde{B}_r=\text{constant}\). The azimuthal component of the momentum equation \eqref{eq:mom} is then
\begin{equation}
    \frac{d}{dr} \Big[ r^3\Big(\rho \tilde{u}_r \tilde{u}_\phi - \frac{\tilde{B}_r \tilde{B}_\phi}{4\pi}\Big) \Big] = 0.
\end{equation}

Next, we apply the Reynolds averaging procedure to the above equation by substituting the Reynolds decomposition \(\tilde{\bm{u}}=\bm{u} + \bm{u}'\) and \(\tilde{\bm{B}}=\bm{B} + \bm{B}'\) and applying the Reynolds averaging operator \(\langle\cdot\rangle\) to each term \citep[][]{Tennekes1972book,usmanov2014three}. The Reynolds average is formally associated with an ensemble average, so that \(\bm{u}=\langle \tilde{\bm{u}}\rangle\) and \(\bm{B}=\langle \tilde{\bm{B}}\rangle\) are the mean velocity and magnetic fields, while \(\bm{u}'\) and \(\bm{B}'\) are fluctuating fields (with arbitrary amplitude). By construction \(\langle\bm{u}'\rangle = \langle \bm{B}'\rangle=0\). We have neglected density fluctuations since observations indicate that solar wind turbulence is nearly incompressible \citep[][]{matthaeus1990JGR}. The Reynolds-averaged azimuthal momentum equation is then
\begin{equation}
    \frac{d}{dr} \Big[ r^3\Big(\rho u_r u_\phi - \frac{B_r B_\phi}{4\pi} + \Rrp \Big) \Big] = 0, \label{eq:Reyn_mom_uphi}
\end{equation}
where 
\begin{equation}
\Rrp = \Big\langle \rho u'_r u'_\phi - \frac{B'_r B'_\phi}{4\pi}\Big\rangle \label{eq:Rrp}
\end{equation}
is the \(r,\phi\) component of the Reynolds stress tensor.  
In deriving Eq. \eqref{eq:Reyn_mom_uphi} we have used standard properties of the Reynolds averaging operation, wherein terms of type \(\langle u_r u_\phi\rangle = u_r u_\phi\), while terms of type \(\langle u_r u'_\phi\rangle = u_r\langle u'_\phi\rangle = 0\) \citep[e.g.,][]{Tennekes1972book}. Since we neglect density fluctuations, \(\rho\) is invariant under the \(\langle\cdot\rangle\) operator.

Eq. \eqref{eq:Reyn_mom_uphi} can be integrated to yield
\begin{equation}
    r^3\left(\rho u_r u_\phi - \frac{B_r B_\phi}{4\pi} + \mathcal{R}_{r\phi}  \right)  = C, \label{eq:azim_intg}
\end{equation}
where \(C\) is a constant independent of \(r\). 
%
%
Reynolds averaging Eq. \eqref{eq:rho} and the solenoidal equation for \(\tilde{\bm{B}}\) yields the two constants \(r^2\rho u_r\) and \(r^2 B_r\). Dividing each term of Eq. \eqref{eq:azim_intg} by \(r^2\rho u_r\) we obtain
\begin{equation}
      \mathscr{L}_T = r u_\phi - \frac{rB_r B_\phi}{4\pi\rho u_r} + \frac{r \mathcal{R}_{r\phi}}{\rho u_r}, \label{eq:L_turb}
\end{equation}
in terms of a new constant \(\mathscr{L}_T  = C/(\rho u_r r^2)\). Eq. \eqref{eq:L_turb} may be compared with the classic WD67 result, Eq. \eqref{eq:WD_L}. Note that the component of Reynolds stress that contributes to the angular momentum balance is an off-diagonal \textit{shear} stress.\footnote{The \(\mathcal{R}_{r\phi}\) component of the Reynolds stress also plays a central role in accretion disk dynamics \citep[e.g.,][]{Balbus2003ARAA}.} 

Applying the Reynolds averaging procedure to Eq. \eqref{eq:ind} yields \(\nabla \times (\bm{u}\times \bm{B} + \bm{\varepsilon}_m\sqrt{4\pi\rho})=0\), where \(\bm{\varepsilon}_m=\langle \bm{u}' \times \bm{B}'\rangle/ \sqrt{4\pi\rho}\) is the mean turbulent electric field \citep{krause1980,breech2003JGR,usmanov2014three}. We neglect \(\bm{\varepsilon}_m\), leaving consideration of its effect on angular momentum to future work. Then the \(\phi\) component of the Reynolds-averaged induction equation above can be integrated to find (following WD67)
\begin{equation}
      r(u_r B_\phi - u_\phi B_r) = -r_0^2 \Omega B_{r,0}, \label{eq:induc}
\end{equation}
where the subscript `\(0\)' denotes the coronal base. 
Substituting \(r^2 B_r = r_0^2 B_{r,0}\) yields
\begin{equation}
   \frac{B_\phi}{B_r} = \frac{u_\phi-r\Omega}{u_r}.
\end{equation}

Using the above in Eq. \eqref{eq:L_turb} to eliminate \(B_\phi\) we find
\begin{equation}
    u_\phi = \frac{\mathscr{L}_T u_r^2/r - r\Omega V_A^2 - \Rrp u_r/\rho}{u_r^2 - V_A^2},
\end{equation}
where \(V_A\equiv V_{A,r} = B_r/\sqrt{4\pi\rho}\) is the radial Alfv\'en speed. Again following WD67, we note that the above equation has a singular point at the Alfv\'en radius where \(u_r = V_A\) and the denominator is zero. To maintain a finite \(u_\phi\) we require the numerator to vanish at \(r=r_A\), giving 
\begin{equation}
    \mathscr{L}_T = r^2_A\Omega(1+\delta_T)
    \label{eq:L_turb2}
\end{equation}
with
\begin{equation}
    \delta_T\equiv \frac{1}{r_A\Omega} \left(\frac{\Rrp/\rho}{V_A} \right)_{r=r_A} = \frac{(4\pi)^{1/2}}{r_A\Omega}  \left(\frac{\Rrp}{B_r \rho^{1/2}} \right)_{r=r_A}\label{eq:delta_T}
\end{equation}

Eq. \eqref{eq:L_turb2} above can be considered a turbulent modification \citep[cf.][]{Hollweg1973JGR} to the classic WD67 result in Eq. \eqref{eq:WD_L_rigid}. 
%
%
Evidently, this modification is inversely proportional to the Alfv\'en radius, the solar rotation rate, the Alfv\'en speed (and equivalently to magnetic field and density) at \(r_A\), in addition to being proportional to the strength of the \(r,\phi\) component of the Reynolds stress at \(r=r_A\). The sign of the turbulent contribution to the angular momentum of the solar wind depends on the sign of \(\Rrp\), 
which means that the relative strength of the velocity and magnetic fluctuations \citep[often called the {Alfv\'en ratio} or the {residual energy}; e.g.,][]{bruno2013LRSP} is a factor in determining the sign of \(\delta_T\).


As usual \citep[see, e.g.,][]{Vidotto2021LRSP}, integrating the angular momentum flux \(\mathscr{L}_T \rho\bm{u}\) over a spherical Alfv\'en surface yields the loss rate of angular momentum including turbulence [cf. Eq. \eqref{eq:WD_Jdot}]:
\begin{equation}
\dot{\mathscr{J}}_{\odot,T} = \frac{2}{3}\Omega r_A^2 \dot{M}_\odot (1+\delta_T),
   \label{eq:J_turb}
\end{equation}
where \(\delta_T\) is assumed spherically symmetric.

\section{Model and Observation based Estimates of the Turbulent Modification to Angular Momentum}\label{sec:results}

\begin{figure*}
    \centering
       \includegraphics[width=.45\textwidth]{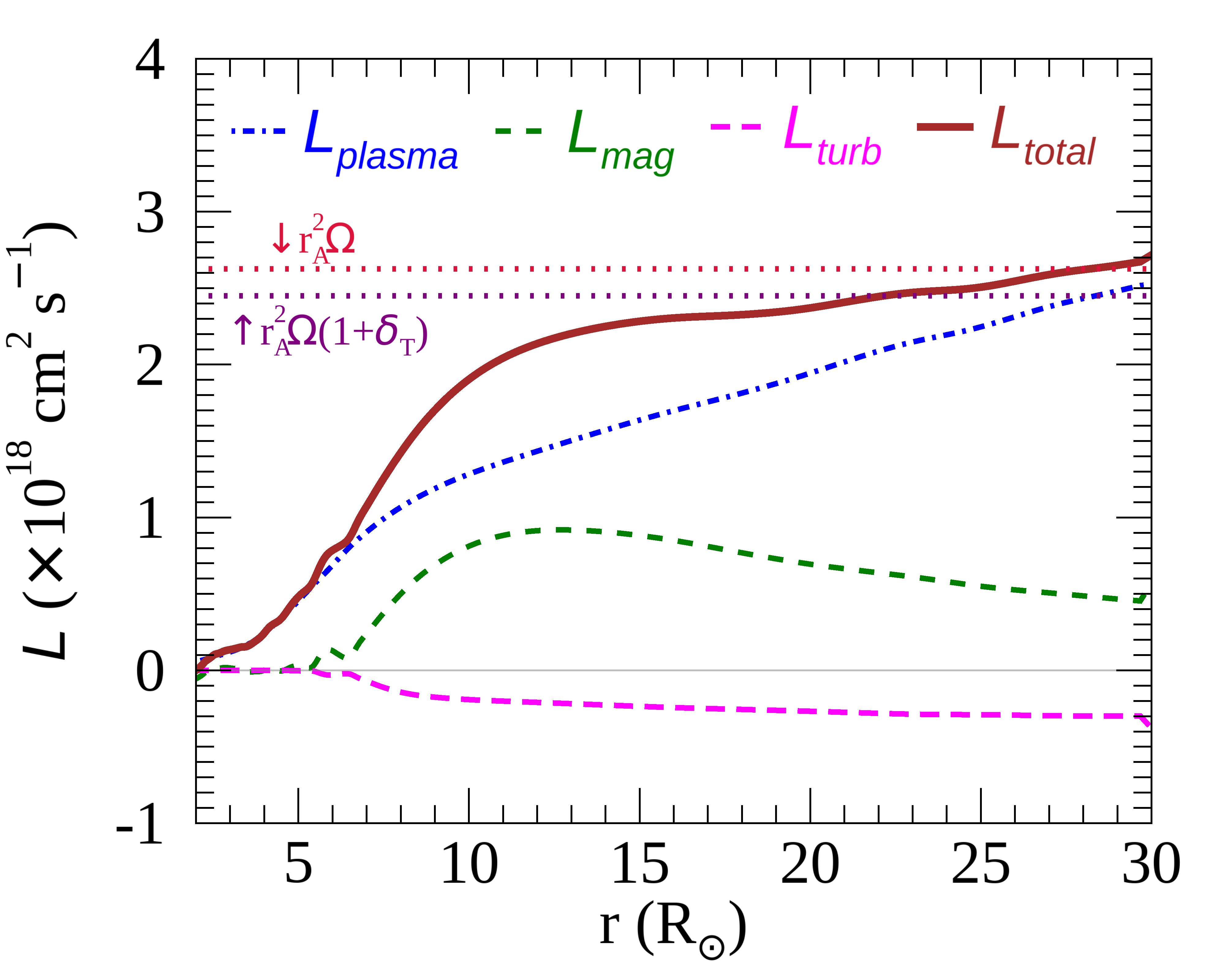}
       \includegraphics[width=.45\textwidth]{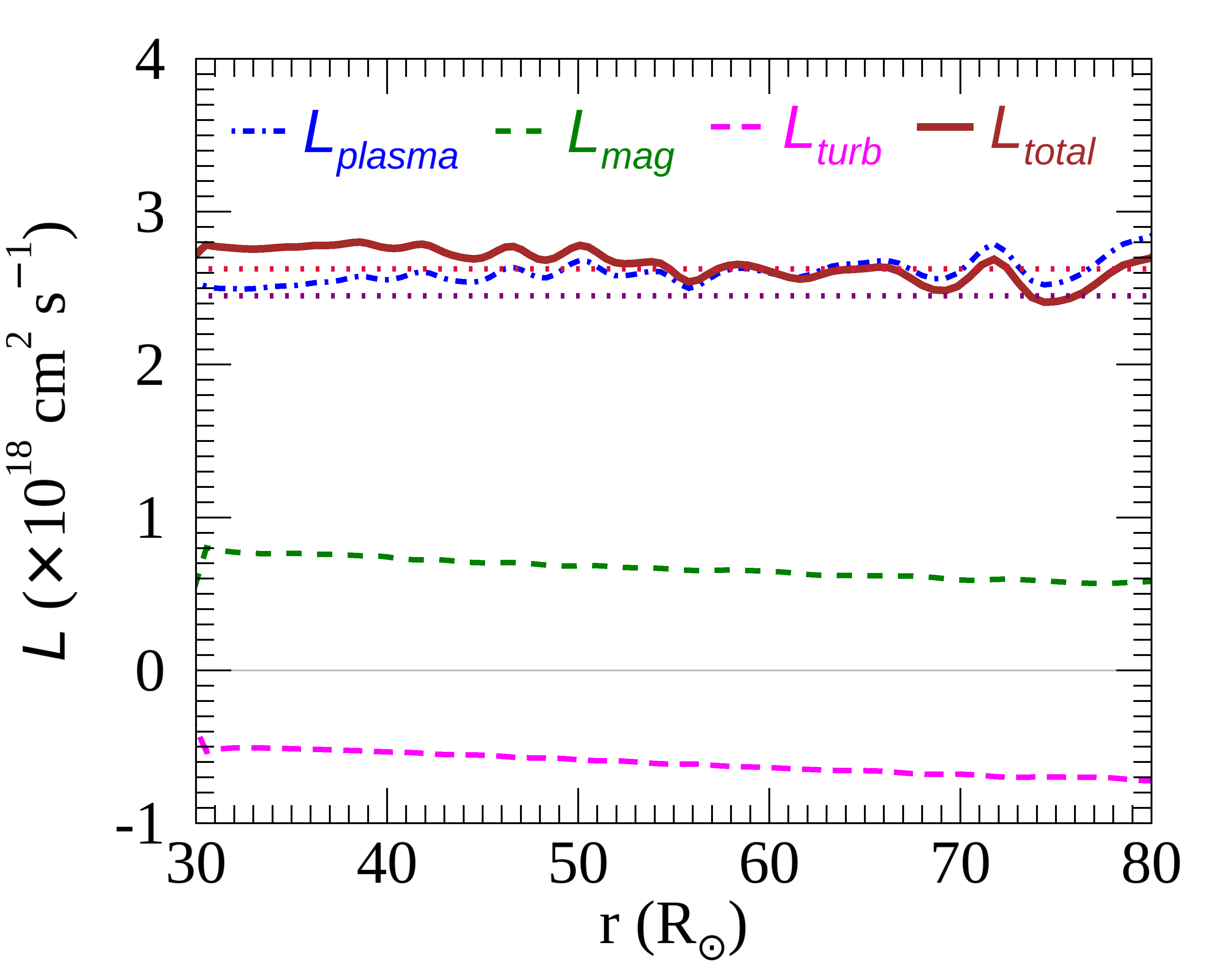}
    \caption{Angular momentum of the solar wind as a function of heliocentric distance \(r\), computed from the Usmanov et al. global model in the solar equatorial plane, as described in text. Here \(\mathscr{L}_\text{total}\) (brown curve) is the sum of \(\mathscr{L}_\text{plasma}=ru_\phi\) (dash-dotted blue curve), \(\mathscr{L}_\text{mag}=-r B_r B_\phi/[4\pi\rho u_r]\) (green curve with short dashes), and \(\mathscr{L}_\text{turb}=r \Rrp/[\rho u_r]\) (magenta curve with long dashes), following Eqns.  \eqref{eq:L_turb} and \eqref{eq:Rrp_closure1}. Slight discontinuity between the values at \(30~R_\odot\) in the left and right panels is due to division of the computational domain into a coronal and a solar-wind section (see Appendix \ref{sec:mod}). 
    Red horizontal dotted line shows the WD67 result \(\mathscr{L}= r_A^2\Omega\); purple horizontal dotted line shows our modified result that includes the effect of turbulence [\(\mathscr{L}_T= r^2_A (1+\delta_T)\); Eq. \eqref{eq:L_turb2}]. The two previous results use \(r_A=13.5~R_\odot\), from our simulation, where \(\delta_T\) is \(-0.07\).} \label{fig:model_r}
\end{figure*}

This section derives quantitative estimates of the turbulent contribution to the solar wind's angular momentum, and the related effects on solar (and stellar) angular momentum loss. 

\vspace{2mm}
\noindent \textit{I}. First, we employ results from the Usmanov et al.  global MHD model of the corona and solar wind \citep{usmanov2011solar,usmanov2012three,usmanov2014three,usmanov2018} to quantify \(\delta_T\). Global heliospheric models are unable to explicitly resolve the \(\bm{u}'\) and \(\bm{B}'\) fluctuations due to computational constraints \citep[][]{miesch2015SSR194,gombosi2018LRSP} and therefore cannot explicitly evaluate the required Reynolds stress \(\Rrp\). This deficiency is alleviated in the Usmanov et al. model by adopting a closure approximation. A key assumption is that the fluctuations are polarized transverse to the mean magnetic field \(\bm{B}\) and axisymmetric about \(\bm{B}\), which is supported by observations of solar wind turbulence \citep[][]{belcher1971JGR,padhye2001JGR,bruno2013LRSP,oughton2015philtran,chhiber2022ApJ}. In this closure, the Reynolds stress tensor reduces to \citep{Usmanov2009ASPC,usmanov2011solar,usmanov2012three,usmanov2016four}
\begin{equation}
    \bm{\mathcal{R}} = \rho K_R(\bm{I} - \hat{\bm{B}}\hat{\bm{B}}),
    \label{eq:Re_closure1}
\end{equation}
where \(\bm{I}\) is an identity matrix, \(\hat{\bm{B}}\) is a unit vector in the direction of \(\bm{B}\), and \(K_R = (\langle u'^2\rangle - \langle b'^2\rangle)/2 = \sigma_D Z^2/2\). Here \(\sigma_D = (\langle u'^2\rangle - \langle b'^2\rangle)/Z^2\) is the normalized energy difference (or residual energy) and \(Z^2=\langle u'^2\rangle + \langle b'^2\rangle\) is twice the fluctuating energy per unit mass, with \(\bm{b}' = \bm{B}' (4\pi\rho)^{-1/2}\). Our model solves a dynamical equation for \(Z^2\) \citep{usmanov2018} and assumes that \(\sigma_D\) remains constant, equal to \(-1/3\). The latter approximation is well supported by observations extending from 0.1 au \citep{chen2020ApJS,parashar2020ApJS,Alberti2022ApJ,Adhikari2024ApJ} to 1 au \citep{bruno2013LRSP}.\footnote{Some new global solar wind models relax the \(\sigma_D=-1/3\) assumption in favor of a dynamical equation for \(\sigma_D\) \citep{Kleimann2023ApJ, usmanov2023AGU}. This will be explored in future work, along with other closures for \(\Rrp\) such as the eddy viscosity approximation \citep{usmanov2018}. Note that \(\sigma_D=-1/3\) corresponds to an Alfven ratio \(\langle u'^2\rangle/\langle b'^2\rangle = 0.5\).} Note that for ``pure'' Alfv\'en waves \(\sigma_D=0\), and the Reynolds stress vanishes. However, solar wind turbulence is essentially never observed to be in such a pristine Alfv\'enic state \citep[e.g.,][]{matthaeus2011SSR,parashar2020ApJS}.

The simulation run used here employs a solar magnetic dipole that is untilted relative to the solar rotation axis. It has been used in a number of previous studies and validated by comparison with observations \citep{usmanov2018,chhiber2019psp1,chhiber2019psp2,chhiber2021ApJ_flrw,chhiber2021AA}. Note that the untilted-dipole simulation has the property of axial symmetry. Some numerical details can be found in Appendix \ref{sec:mod}.\footnote{Strictly speaking, the use of a 3D simulation with dipolar magnetic structure to evaluate Eq. \eqref{eq:L_turb} and \eqref{eq:L_turb2} is not a  self-consistent approach, since this type of simulation may have non-zero \(B_\theta\) that is normal to the solar equatorial plane, especially in the helmet streamer region in the low corona \citep[see, e.g.,][]{usmanov2000global}. In contrast, the WD67-style model developed in Sec. \ref{sec:theory} neglects \(B_\theta\). Nevertheless, we choose to employ the Usmanov et al. model here since our goal is to estimate \(\delta_T\), which requires an evaluation of the Reynolds stress tensor. We are not aware of any other solar wind model that accounts for this latter quantity. 
}

From Eq. \eqref{eq:Re_closure1} we have 
\begin{equation}
    \Rrp = -\rho\sigma_D Z^2 \frac{B_r B_\phi}{2 B^2},
    \label{eq:Rrp_closure1}
\end{equation}
where \(B^2 = B^2_r + B^2_\phi + B^2_\theta\). With this, we compute each term on the r.h.s. of Eq. \eqref{eq:L_turb} from the simulation data at a heliolatitude of \(\sim 2\degree\), chosen to be very near the equatorial plane but not immediately within the heliospheric current sheet (HCS) of the model where \(B_r\) vanishes \citep{usmanov2018,chhiber2019psp1}. Results are shown in Fig. \ref{fig:model_r} as a function of heliocentric distance \(r\). Also shown are dotted horizontal lines indicating the value of \(\mathscr{L}\) computed from Eqs. \eqref{eq:WD_L_rigid} and  \eqref{eq:L_turb2}, using the simulation Alfv\'en radius at \(2\degree\) latitude, found to be \(13.5~R_\odot\). \(\Omega\) is taken to be the sidereal rotation rate at the equator, \(2.972 ~\mu \text{~rad s}^{-1}\) \citep{Snodgrass1990ApJ}. 

Fig. \ref{fig:model_r} shows that the total \(\mathscr{L}\) rapidly increases up to \(\sim 10~R_\odot\) after which it asymptotes to a near-constant value. The behavior at \(r \lesssim 10~R_\odot \) is a consequence of the closed loop (helmet streamer) topology of the magnetic field near the Sun \citep{gombosi2018LRSP}, where \(B_r\) is very small near the equator and \(B_\theta\) is finite, which violates an assumption used to derive Eq. \eqref{eq:L_turb}. We note that the dominant contribution comes from the plasma, while magnetic and turbulent contributions are similar in magnitude. The dominance of the plasma term over the magnetic term is again due to the proximity of the HCS to the equator. As one moves to slightly larger heliolatitudes the magnetic term becomes the dominant contribution, and the corresponding plot appears consistent with the result shown in WD67 \citep[see also][]{usmanov2018}. This 3D variation and global structure is examined in detail in a companion paper (Chhiber et al. 2025, in prep.)

If the turbulent contribution were to be neglected, then the total (near constant) \(\mathscr{L}\) would be computed as \(\sim 3\times 10^8 ~\text{cm}^2 ~\text{s}^{-1}\). The turbulent contribution therefore reduces the total by \(\sim 10\%\), with a similar effect on \(\mathscr{\dot{J}}_T\). The crucial role played by the \(\sigma_D\) parameter in determining the sign of the turbulent contribution should be emphasized here. In the coronal and inner heliospheric solar wind \(\sigma_D\) tends to be negative, corresponding to magnetically-dominated turbulence. This evidently provides a negative contribution to \(\mathscr{L}_T\), and equivalently, a positive contribution to \(u_\phi\) [as can be seen by inverting Eq. \eqref{eq:L_turb} to solve for \(u_\phi\)], implying a boost to corotation of the solar wind. This effect of the turbulence can be interpreted an action of a \textit{positive} viscosity \citep[see also][]{Weber1970JGR}. If \(\sigma_D\) is positive  (kinetically-dominated turbulence, possibly associated with vortical structures) then the turbulent contribution to to \(\mathscr{L}_T\) would be positive, implying a tendency to reduce corotation, which can be interpreted as a negative viscosity  \citep[cf.][]{Hollweg1973JGR}. Very small positive values of \(\sigma_D\), while rare, have been observed in the solar wind \citep[e.g.,][]{wicks2013apj}.

\begin{figure}
    \centering
     \includegraphics[width=.49\textwidth]{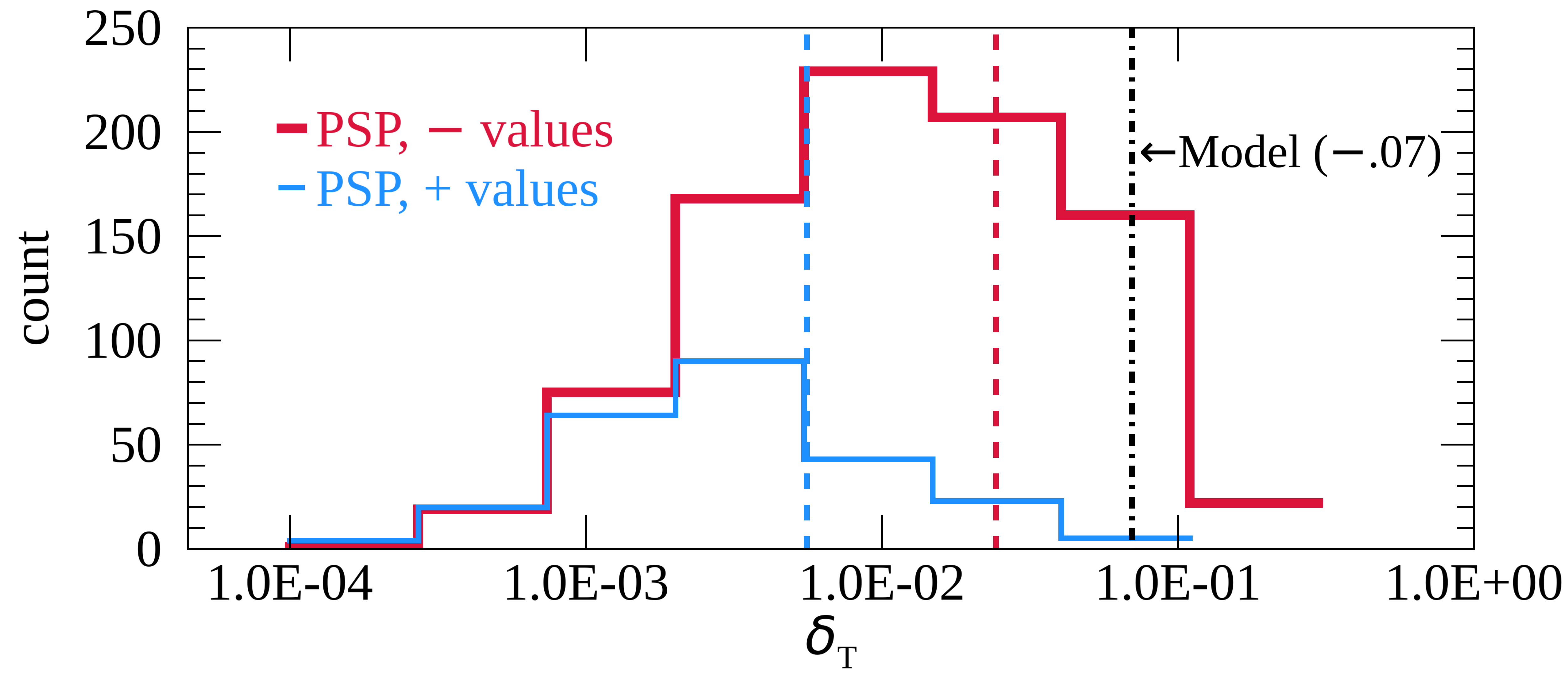}
    \caption{Solid curves show histograms of \(\delta_T\), the turbulent modification to the WD67 angular momentum [Eq. \eqref{eq:delta_T}], computed from PSP observations at the Alfv\'en surface, aggregated during solar encounters 8, 9, and 10. \(\Rrp\) is computed directly from Eq. \eqref{eq:Rrp} (see text for details). Thick red (thin blue) curve corresponds to negative (positive) values of \(\delta_T\), while dashed red and blue vertical lines mark the mean values for the two cases. Dash-dotted black vertical line marks the value obtained from the Usmanov et al. model, as described in the text, using the closure approximation Eq. \eqref{eq:Rrp_closure1} for \(\Rrp\).} \label{fig:delta_hist}
\end{figure}
\vspace{2mm}

\noindent \textit{II}. Next, we employ PSP measurements at the solar wind's Alfv\'en surface to estimate \(\delta_T\), with a \textit{direct} computation of \(\Rrp\) [Eq. \eqref{eq:Rrp}] from the high-resolution observations. Details of data selection and initial processing are in Appendix \ref{sec:psp}. Estimation of \(\delta_T\) requires computing both mean and fluctuating quantities [Eq. \eqref{eq:delta_T}]. We compute mean fields by applying a boxcar average to the time series of \(\tilde{\bm{u}},~\tilde{\bm{B}}\), and number density, over a moving window of 600-s duration (60 points in the 10-s cadence time series). This corresponds to averaging over a few correlation times, a standard approach in turbulence studies to separate mean and fluctuating fields \citep[e.g.,][]{isaacs2015JGR120,chhiber2021ApJ_psp}. Thus we obtain time series of mean fields (\(\bm{u},~\bm{B},~\rho\)), where \(\rho=m_p n\) with proton mass \(m_p\). Time series of fluctuating fields are computed in the usual way as \(\bm{u}'=\tilde{\bm{u}} - \bm{u}\) and \(\bm{B}'=\tilde{\bm{B}} - \bm{B}\). Applying the 600-s moving boxcar average to \(\rho u'_r u'_\phi\) and \(B'_r B'_\phi\) immediately gives us the time series of \(\Rrp\) [Eq. \eqref{eq:Rrp}]. The radial component of the Alfv\'en velocity is computed as \(V_{A,r} \equiv V_A = B_r/\sqrt{4\pi\rho}\).

The final step is to identify Alfv\'en surface crossings, for which we compute the Alfv\'en Mach number \(M_A= u_r/V_A\), and identify points where \(0.9\le M_A \le 1.1\), which includes data at or very close to the Alfv\'en surface. Note that PSP is moving extremely fast here \citep{fox2016SSR} and allowing for this range of \(M_A\) gives us a larger sample compared to only considering the very brief instants that PSP crosses the Alfv\'en surface. \(\delta_T\) is then computed at these points using Eq. \eqref{eq:delta_T}.

The resulting distribution of \(\delta_T\) is shown in Fig. \ref{fig:delta_hist}. Observed values evidently show large variability, as is commonly the case for turbulence parameters in the solar wind \citep[e.g.,][]{bruno2013LRSP,isaacs2015JGR120}. Negative \(\delta_T\) values (mean \(\sim -0.03\)) are much more likely than positive ones, so the overall qualitative trend is consistent with the model.

\vspace{2mm}

\noindent \textit{III}. Finally, we carry out a simple analysis to get a rough quantitative estimate of the effect of turbulence on the long-term evolution of the solar rotation rate. Our analysis is based on the model for evolution of stellar rotation in the main sequence described by \cite{Vidotto2021LRSP}, who derives analytically the well-known empirical relation \(\Omega \propto t^{-1/2}\) of \cite{Skumanich1972ApJ}. We emphasize that the stellar evolution model used here is highly simplified, and the purpose is to quantify the effect of turbulence on stellar rotational evolution in the absence of other, possibly more significant factors. For example, a more realistic study should not assume a constant mass-loss rate since relaxing this assumption \citep{Wood2002ApJ,Holzwarth2007AA} could produce a lower-order effect than that of turbulence.\footnote{The mass-loss rate, in turn, can itself be influenced by turbulence amplitudes at the coronal base \citep{airapetian2016reconstructing}}. With this caveat in mind, we start with the WD67 result [Eq. \eqref{eq:WD_Jdot}] for the loss-rate of the Sun's angular momentum:
\begin{equation}
    \dot{\mathscr{J}}_\odot = -\frac{2}{3}\frac{\Omega}{V_{A,A}}(B_{r,\odot} R_\odot^2)^2, \label{eq:Jdot_1}
\end{equation}
where we have used \(\dot{M}_\odot = 4\pi r_A^2 \rho_A V_{A,A}\) for the constant mass-loss rate, and the conservation of magnetic flux \(B_{r,\odot}R_\odot^2 = B_{r,A}r_A^2\). Here subscripts \(\odot\) and \(A\) refer to the solar surface and the Alfv\'en surface, respectively. The `\(-\)' sign appears since the rate of change of angular momentum of the Sun is negative, due to the angular momentum gained by the solar wind.

Assuming the Sun to be a uniform spherical solid body with moment of inertia \(\mathcal{I}\), its angular momentum is \(\mathscr{J}_\odot = \mathcal{I}\Omega = \frac{2}{5} M_\odot R_\odot^2 \Omega\), which can also be used to define 
\begin{equation}
   \dot{\mathscr{J}}_\odot = \frac{2}{5} M_\odot R_\odot^2 \frac{d\Omega}{dt}.
\end{equation}
We have assumed that \(R_\odot\) is independent of age and that change in solar mass is small compared to the change in \(\Omega\). Equating the above to \eqref{eq:Jdot_1} we get
\begin{equation}
    dt = -\frac{3}{5} \frac{M_\odot}{R_\odot^2}\frac{V_{A,A}}{B_{r,\odot}^2} \frac{d\Omega}{\Omega}.
\end{equation}
This equation must be integrated to obtain \(\Omega (t)\), which describes the rotational evolution of the Sun. We assume that \(B_{r,\odot}\) is linearly dependent on \(\Omega\) \citep[i.e., a linear-type dynamo;][]{Durney1982ApJ}, which is roughly consistent with observations \citep{Vidotto2014MNRAS}. \(V_{A,A}\) depends on the magnetic field and density at the Alfv\'en point, and for simplicity it is assumed that this velocity is independent of \(\Omega\). We then have
\begin{equation}
    \int dt = -\frac{3}{5} \frac{M_\odot V_{A,A}}{R_\odot^2} \int \frac{1}{B_{r,\odot}^2} \frac{d\Omega}{\Omega} = C_1 \int \frac{1}{\Omega^3}  d\Omega, 
\end{equation}
which can be integrated to get \(\Omega\propto t^{-1/2}\). Constant factors have been gathered in \(C_1\). Carrying out the integration between two reference times \(t_1\) and \(t_2\) yields the equation
\begin{equation}
    t_2 - t_1 = -\frac{C_1}{2} \bigg( \frac{1}{\Omega_2^2} - \frac{1}{\Omega_1^2} \bigg).    \label{eq:Omega_t}
\end{equation}

\begin{figure}
    \centering
       \includegraphics[width=.45\textwidth]{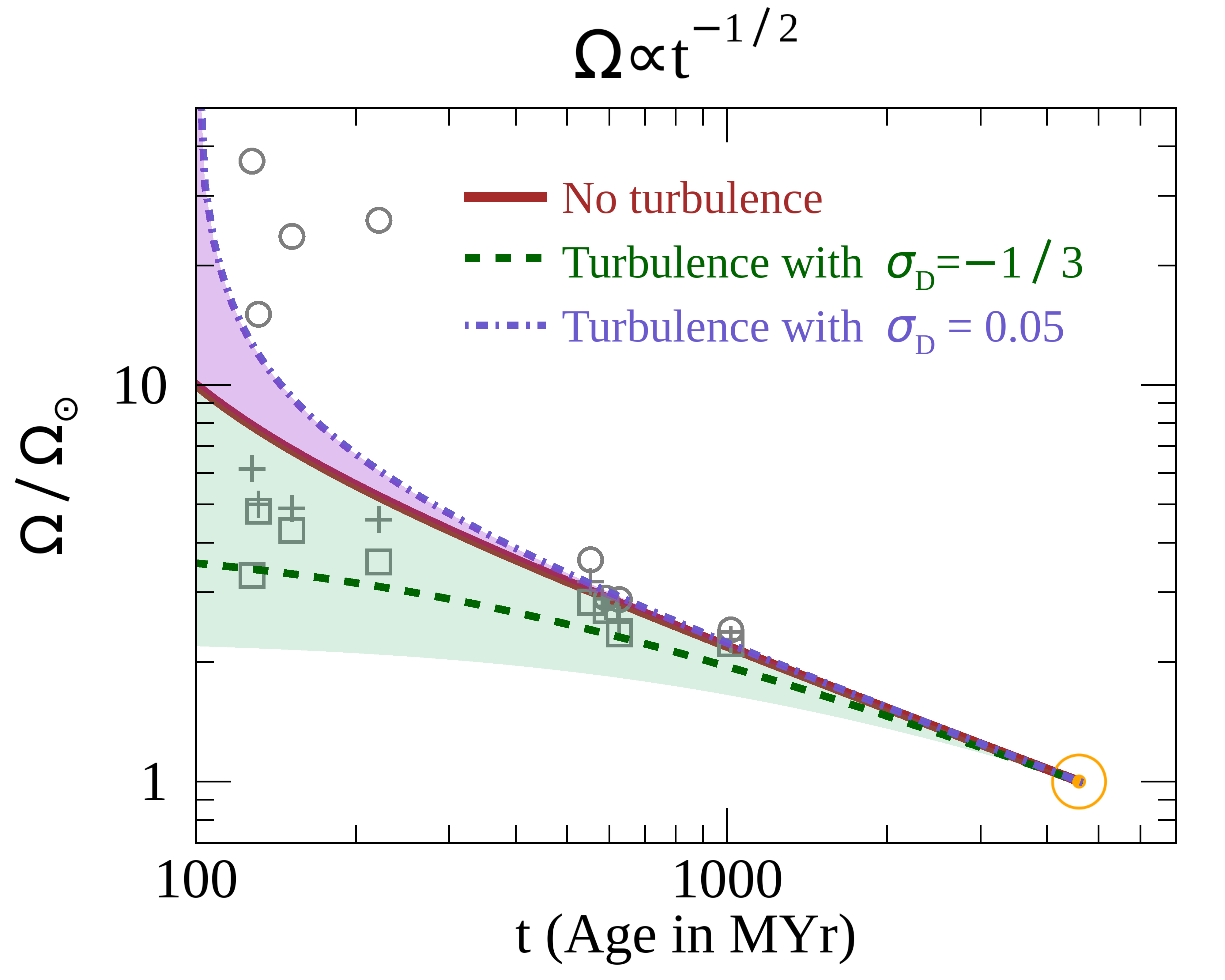}
    \caption{Evolution of stellar rotation rate \(\Omega\) with age. \(\Omega\) is normalized by the present-day rotation rate of the Sun, \(\Omega_\odot\). Symbols show observations of rotation rates of solar-mass stars taken from \cite{Gallet2015AA}: \(\square,~+,\) and \(\circ\) symbols represent 25th, 50th, and 90th percentiles, respectively, of the observations at a particular age. Green dashed curve is based on \(\delta_T=-.07\), the value estimated from the global model (see Fig. \ref{fig:model_r}). Lower and upper bounds of green-shaded region are based on \(\sigma_D\) equal to \(-1\) and 0, respectively, keeping other factors in \(\delta_T\) constant; the former case represents purely magnetic fluctuations while the latter case is that of pure Alfv\'en waves with equal kinetic and magnetic fluctuation energy. Dash-dotted blue curve is based on \(\sigma_D=0.05\), a very slight preponderance of kinetic fluctuation energy. See text for more details.} \label{fig:omega}
\end{figure}

Our next step is to find a solution to the above equation that satisfies \(\Omega_2\) = \(\Omega_\odot\) and \(\Omega_1 = 10 \Omega_\odot\), where the current solar rotation rate \(\Omega_\odot\) is used as a reference normalization factor, \(t_2=4.6 \times 10^3\) MYr is the current age of the Sun, and \(t_1 = 100\) MYr. The chosen values of \(t_1\) and \(\Omega_1\) have been motivated by typical observations of main-sequence stars \citep{Gallet2015AA}. With \(C_1\) thus determined, the solid brown curve in Fig. \ref{fig:omega} shows this solution. Data from \cite{Gallet2015AA} are shown as symbols.

To obtain insight on how turbulence modifies this solution, we note that the factor \(f_T = 1+\delta_T\) in Eq. \eqref{eq:J_turb} can be shown to act as a multiplicative factor to the constant \(C_1\) in Eq. \eqref{eq:Omega_t}. Here we have once again assumed for simplicity that \(f_T\) remains constant, like \(M_\odot\) and \(V_{A,A}\). With this modified \(C_1\), we solve Eq. \eqref{eq:Omega_t} for \(\Omega(t)\) such that \(\Omega(t=4.6\times 10^3 ~\text{MYr})=\Omega_\odot\). The results are shown in Fig. \ref{fig:omega} and represent different possible rotational histories of the Sun, accounting for turbulence: dashed and dash-dotted curves and shaded regions show the modification to \(\Omega(t)\) by variation in \(\sigma_D\) (see caption). It is noteworthy that turbulence characterized by magnetic fluctuation energy greater than velocity fluctuation energy (\(\sigma_D<1\)) exhibits a slower decline of angular momentum. This condition is that which is usually found in the solar wind. On the other hand,  turbulence with greater inertial range velocity-field energy (\(\sigma_D>1\))would cause faster decline of angular momentum in the present formulation. Keeping in mind the caveat mentioned at the start of this sub-section, further work will be needed to better understand the interplay of turbulence with other important factors that govern stellar rotational evolution.

%
\section{Conclusions and Discussion}

We have made an attempt to evaluate the influence of turbulence in the solar wind on its angular momentum, and on the long-term rotational evolution of the Sun. The analytical theory, derived in the Reynolds-averaged MHD framework absent any linearized treatments, expresses the effect of turbulence as modifications to the well-known formulae from \cite{weber1967ApJ148}. Quantitative analyses, employing a numerical model and new PSP observations at the solar wind's Alfv\'en surface, indicate that the contribution of turbulence to the wind's angular momentum is not negligible compared to that due to the bulk plasma flow and magnetic field, and tends to be negative. We also used a simplified model of long-term rotational evolution of Sun-like stars to quantify the impact of turbulence on stellar ``spin-down'' in the main sequence.

Future work can aim to alleviate limitations of our brief initial study. Like WD67, our approach is based on a solution for the equatorial wind with axial symmetry. We examine the effects of 3D structure, non-axisymmetry, and varying solar activity \citep[e.g.,][]{Cohen2014ApJ,reville2015ApJ798,Finley2018ApJ} in a companion paper focused on numerical analyses of 3D global simulations. The crude assumptions of constant wind parameters (e.g., mass-loss rate, moment of inertia) over the solar lifetime can also be refined \citep[e.g.,][]{Kawlaler1988ApJ,Holzwarth2007AA,Reville2016ApJ,Finley2019ApJ_883}, possibly by employing a ``best-fit'' formulation for  \(\dot{\mathscr{J}}\) that accounts for variability in the relevant parameters \citep[e.g.,][]{Matt2012ApJ}. Finally, turbulence is associated with complexity and ``fuzziness'' of the Aflv\'en surface \citep{deforest2018ApJ,Wexler2021ApJ, Chhiber2022MNRAS,Chhiber2024MNRASL}, and it could be worthwhile to consider what impact this may have on solar angular momentum loss.




\vspace{1cm}
RC acknowledges useful discussions with Junxiang Hu and Vladimir Airapetian. PSP data are publicly available at the \href{https://spdf.gsfc.nasa.gov/}{NASA Space Physics Data Facility}. This research is partially supported by NASA under Heliospheric Supporting Research program grants 80NSSC18K1648 and 80NSSC22K1639, and Living With a Star (LWS) Science program grant 80NSSC22K1020. PSP was designed, built, and is now operated by the Johns Hopkins Applied Physics Laboratory as part of NASA’s LWS program (contract NNN06AA01C). This work utilized resources provided by the Delaware Space Observation Center (DSpOC) numerical facility. 

\appendix

\section{Some Numerical Details of the  Solar Wind Model}\label{sec:mod}

The simulation domain extends from the coronal base at \(1~R_\odot\) to 3 au, and is divided into two regions: an inner (coronal) region of \(1\dash 30~R_\odot\)
and an outer (solar wind) region between \(30~R_\odot\) and 3 au. The
relaxation method, i.e., integration of equations in time until a steady state is achieved, is used in both
regions, with the \(30~R_\odot \) boundary of the coronal region providing inner boundary values for the solar wind region. Input parameters specified at the coronal base include: amplitude of fluctuations/Alfv\'en waves \(\sim 30~ \kmps\), density \(\sim 1\times 10^8 ~\text{cm}^{-3} \), temperature \(\sim 1.8\times 10^6~\degree\text{K}\), and solar magnetic dipole strength 12 G. The numerical resolution is \(702\times 120\times 240\) grid points along \(r\times \theta\times \phi\) coordinates, with logarithmic spacing along heliocentric radius (\(r\)) that becomes larger with increasing \(r\). Latitudinal (\(\theta\)) and longitudinal (\(\phi\)) grids have equidistant spacing of \(1.5\degree\)~each. In terms of physical scales, the grid spacing corresponds to roughly a few correlation lengths of magnetic fluctuations \citep[][]{Ruiz2014SoPh,chhiber2021ApJ_psp}. For further numerical details see \cite{usmanov2012three,usmanov2014three,usmanov2018}.

\section{Details of PSP data selection and processing}\label{sec:psp}
Publicly-available data from solar encounters (E) 8 (Apr-May 2021), 9 (Aug 2021), and 10 (Nov 2021) are employed.\footnote{The exact time periods of the encounters are listed on the \href{https://psp-gateway.jhuapl.edu/website/SciencePlanning/MissionEvents}{PSP Science Gateway}.} PSP first observed sub-Alfv\'enic flow in Apr 2021 \citep{kasper2021prl} and several Alfv\'en surface crossings occurred across E8-E10, at \(r\sim 13\dash 20~R_\odot\) \citep{Chhiber2024MNRASL}. Magnetic field measurements by the fluxgate magnetometer on the FIELDS suite \citep{bale2016SSR} are used, while velocity measurements are from the SPAN-I instrument aboard the SWEAP suite \citep{kasper2016SSR,Livi2022ApJ}. The FIELDS dataset also provides electron number density via quasi-thermal noise spectroscopy \citep{Moncuquet2020ApJS}, and we use these data as a proxy for mean proton number density \(n\), justified by the assumption of quasi-neutrality in the solar wind plasma charge. Note that computation of \(\Rrp\) from Eq. \eqref{eq:Rrp} requires only the \textit{mean} mass density \(\rho\), therefore fine variations in this quantity are not relevant to our purpose. 

The data time series were downsampled to 10-s cadence; this allows for sampling of fluctuations at scales corresponding to around a decade of the low frequency (or wavenumber) end of the inertial range of solar wind turbulence \citep{Kiyani2015RSPTA}, since the typical spectral break separating the inertial range from energy-containing scales occurs at \(\sim 100\dash 500\) s at these heliocentric distances \citep{kasper2021prl,Zhao2022ApJ_subA}. A 10-s cadence also puts us well above the kinetic range of scales where the MHD approximation breaks down \citep{goldstein2015RSPTA}; at radial distances of interest this occurs below \(\sim 1\) s \citep{kasper2021prl}.


\bibliography{chhibref}

\end{document}